\documentclass{amsart}
\usepackage{amsfonts, amsmath, amssymb, amsthm, bbm}
\usepackage{calc}
\usepackage[english]{babel}
\usepackage[T1]{fontenc}
\usepackage{graphicx}
\usepackage{hhline}
\usepackage[hidelinks]{hyperref}
\usepackage[utf8]{inputenc}
\usepackage{multirow}
\usepackage{setspace}
\usepackage{textcomp, tabulary, stmaryrd}
\usepackage{verbatim}

\theoremstyle{definition}

\newcommand{\bib}[4]{#1 (#2). #3. \textit{#4}.}

\newcommand{\post}{\text{post}}
\newcommand{\pre}{\text{pre}}

\newcommand{\R}{\mathbb{R}}

\newcommand{\luft}{\vspace{1em}}

\renewcommand{\quote}[1]{\guillemotleft#1\guillemotright}

\allowdisplaybreaks

\begin{document}


\title{Deep Replication of a Runoff Portfolio}

\author{Thomas Krabichler}
\address{}
\curraddr{}
\email{thomas.krabichler@ost.ch}
\thanks{}

\author{Josef Teichmann}
\address{}
\curraddr{}
\email{josef.teichmann@math.ethz.ch}
\thanks{}


\date{}

\begin{abstract}
To the best of our knowledge, the application of deep learning in the field of quantitative risk management is still a relatively recent phenomenon. This article presents the key notions of Deep Asset Liability Management (\quote{Deep~ALM}) for a technological transformation in the management of assets and liabilities along a whole term structure. The approach has a profound impact on a wide range of applications such as optimal decision making for treasurers, optimal procurement of commodities or the optimisation of hydroelectric power plants. As a by-product, intriguing aspects of goal-based investing or Asset Liability Management (ALM) in abstract terms concerning urgent challenges of our society are expected alongside. We illustrate the potential of the approach in a stylised case.
\end{abstract}

\maketitle

\section{Introduction}

Mismatches between opposing fixed and floating prices along a whole term structure are omnipresent in various business fields and raise substantial business risks. Already humble approaches lead to analytically intractable mathematical models entailing high-dimensional allocation problems with constraints in the presence of frictions. This impediment often leads to over-simplified modelling (which still needs sophisticated technology), uncovered risks or overseen opportunities. Spectacular successes of deep learning techniques in language processing, image recognition, learning games from tabula rasa, or risk management, just to name a few, are stimulating our imagination that \emph{asset-liability-management} (ALM) might enter a new era. This novel approach, briefly denoted \quote{Deep~ALM}, raises a huge field of mathematical research questions such as feasibility, robustness, inherent model risk and inclusions of all economic aspects. If we answer these questions, this could possibly change the ALM practice entirely.

ALM is a research topic that may be traced back to the seventies of the previous century when Martin L.~Leibowitz and others developed cash-flow matching schemes in the context of the so-called \emph{dedicated portfolio theory}; see \cite{FLS92}. In the eighties and nineties, ALM was an intensively studied research topic in the mathematical community. This unanimously important branch of quantitative risk management has been dominated by techniques from \emph{stochastic control theory}; e.g., see \cite{CD98} by Giorgio Consigli and Michael~A.~H.~Dempster or \cite{FS03} by Karl~Frauendorfer and Michael~Sch\"urle for articles utilising \emph{dynamic programming}. Further aspects of the historical evolution are treated in \cite{R13} by Ronald~J.~Ryan. However, already over-simplified modelling involves a high complexity both from the analytical and the technological viewpoint. A cornerstone in this regard is \cite{KZ83} by Martin~I.~Kusy and William~T.~Ziemba. They consider a rather advanced and flexible ALM model by specifying an economic objective (e.g., the maximisation of the present value of future profits), a series of constraints (e.g., regulatory requirements and liquidity assurance) and penalty costs for constraint violations.

Prevalent approaches applied in the industry such as solely immunising oneself against parallel shifts of the yield curve have remained suboptimal until this day; e.g., see \cite{SDR19} by Martin~Spillmann, Karsten~D\"ohnert and Roger~Rissi for a recent exposition. Particularly, the empirical approach with static replicating portfolios described in the Section~4.2 of \cite{SDR19} is often applied in mid-sized retail banks. Although the impact of stress scenarios is investigated in the light of regulatory requirements, the holistic investment and risk management process of many treasury departments remains insufficient. More sophisticated approaches, such as for instance \emph{option-adjusted spread models} (e.g., see \cite{Ba07} by Martin~M.~Bardenhewer), have not become established in the financial industry. Therefore, despite many achievements, it is fair to say that ALM has not entirely fulfilled its high expectations.

The fresh approach on \emph{deep hedging} by Hans~B\"uhler, Lukas~Gonon, Josef~Teichmann and Ben~Wood (see \cite{BGTW19}) might pave a way for a new era in quantitative risk management. It is clear that classical yet analytically intractable problems from ALM can be tackled with techniques inspired from deep reinforcement learning in case of an underlying Markovian structure; see \cite{GBC16} by Ian~Goodfellow, Yoshua~Bengio and Aaron~Courville for a comprehensive overview of deep learning and \cite{SB98} by Richard~S.~Sutton and Andrew~G.~Barto for an introduction to reinforcement learning in particular. Reinforcement learning has found many applications in games (e.g., see \cite{DM18} by a research team of Google's DeepMind), robotics (e.g., see \cite{KBP13} by Jens~Kober, J.~Andrew~Bagnell and Jan~Peters) and autonomous vehicles (e.g., see \cite{V20} by a joint publication of Valeo and academic partners). Recent monographs on machine learning for financial applications are \cite{LDP18} by Marcos L\'opez de Prado and \cite{DHB20} by Matthew~F.~Dixon, Igor~Halperin and Paul~Bilokon. Regarding applications of neural networks for hedging and pricing purposes, some $200$ research articles have accumulated over the last 30~years; \cite{RW19} by Johannes~Ruf and Weiguan~Wang provides a comprehensive literature review. Only a handful of these articles exploit the reinforcement learning paradigm. In contrast to this obvious applications of deep reinforcement learning to ALM, \emph{deep hedging approaches to ALM} are simpler but still providing the solution
relevant for business decisions. Notice, however, that neither Markovian assumptions are needed, nor value functions or dynamic programming principles: Deep~ALM will simply provide an artificial asset liability manager who precisely solves the business problem (and not more) in a convincing way, i.e.~provides ALM strategies along pre-defined future scenarios and stress scenarios. The fundamental ideas will be further elaborated on in the sections below. To the best of our knowledge, we are the first to tackle ALM systematically by deep hedging approaches.

There are many fundamental problems arising in mathematical finance that cannot be treated analytically. Exemplarily, explicit pricing formulae for American options in continuous time are yet unknown even in the simplest settings such as the Black-Scholes-framework. This is incredible in that most exchange traded options are of American style. Optimal stopping problems in genuine stochastic models involve so-called parabolic integro-differential inequalities, which can be numerically approximated by classical methods up to three dimensions; e.g., see the Section~10.7 in \cite{HRSW13}. The curse of dimension impedes the feasibility of classical finite element methods for applications involving more than three risk drivers. With today's computational power of a multi-core laptop, the runtime in six dimensions would last many centuries. A popular circumvention is the regression based least squares Monte-Carlo method proposed by Francis A. Longstaff and Eduardo S. Schwartz; see \cite{LS01}. Some statisticians perceive machine learning as a generalisation of the regression technique. In this sense, a spectacular success was achieved by Sebastian Becker, Patrick Cheridito and Arnulf Jentzen when they utilised deep neural networks to price American options in as much as $500$ dimensions below $10$~minutes; see \cite{BCJ19} for further details. One is forced to abandon the fundamental desire of directly solving a very complex mathematical equation and must approve of the paradigm shift to the deep learning principle: \quote{What would a clever and non-forgetful artificial financial agent with a lot of experience and a decent risk appetite do?}. This change of mindset opens the room for many possibilities not only in quantitative finance. Computationally very intensive techniques such as nested Monte-Carlo become obsolete. We aim at utilising these advances to tackle problems that were formulated back in the eighties, e.g., by Martin I. Kusy and William T. Ziemba (see \cite{KZ83}), and have remained unsolved for the desired degree of complexity ever since. Thus, the technological impact of Deep~ALM for the financial industry might be considerable.

The Markov assumption, roughly speaking that future states only depend on the current state and not the past, usually leads to comparatively tractable financial models. Real world financial time series often feature a \emph{leverage effect} (i.e., a relationship between the spot and the volatility) and a \emph{clustering effect} (i.e., a persistence of low and high volatility regimes). Unless volatility becomes a directly traded product itself (which generally is not the case) and is perceived as an integrable part of the state, an effective hedging strategy can hardly be found analytically. This leads to the concept of \emph{platonic financial markets} that was introduced in \cite{CKT17} by Christa~Cuchiero, Irene~Klein and Josef~Teichmann. Generally observable events and decision processes are adapted to a strict subfiltration of its counterpart that is generated by the obscure market dynamics. This concept goes in a similar direction as \emph{hidden Markov models}; e.g., see the textbook \cite{EAM95} by Robert~J.~Elliott, Lakhdar Aggoun and John~B.~Moore for a reference. Regarding Deep~ALM, this inherent intractability of platonic financial markets is not an impediment at all. Certainly, the replication portfolios along a trained deep neural network only depend on generally observable states. However, the financial market dynamics do not have to satisfy any Markovianity restrictions with respect to the observable states and/or traded products. The deep neural network jointly learns the pricing and the adaptation of hedging strategies in the presence of \emph{market incompleteness} and \emph{model uncertainty}; e.g. see \cite{EKQ95} by Nicole~El~Karoui and Marie-Claire~Quenez and \cite{ALP95} by Marco~Avellaneda, Arnon~Levy and Antonio~Par\'as for further details. Hence, Deep~ALM unifies key elements of mathematical finance in a fundamental use case.

An essential prerequisite for the viability of Deep ALM in a treasury department is to come up with a sufficiently rich idea of the macro-economic environment and bank-specific quantities such as market risk factors, future deposit evolutions, credit rate evolutions and migrations, stress scenarios and all the parameterisations thereof. In the end, the current state of a bank must be observable. In contrast to dynamic programming, machine learning solutions are straightforward to implement directly on a cash-flow basis without any simplifications (e.g., a divisibility into simpler subproblems becomes redundant), and their quality can be assessed instantaneously along an arbitrarily large set of validation scenarios. Moreover, the paradigm of letting a smart artificial financial agent (similar to an artificial chess player) with a lot of experience choose the best option allows to tame the curse of dimension, which is the pivotal bottleneck of classical methods. Further aspects such as intricate price dynamics, illiquidity, price impacts, storage cost, transaction cost and uncertainty can easily be incorporated without further ado; all these aspects are hardly treated in existing frameworks. Despite the higher degree of reality, Deep~ALM is computationally less intensive than traditional methods. Risk aversion or risk appetite respectively can be controlled directly by choosing an appropriate objective associated with positive as well as negative rewards in the learning algorithm. Regulatory constraints can be enforced through adequately chosen penalties. Hence, Deep~ALM offers a powerful and high-dimensional framework that supports well-balanced risk-taking and unprecedented risk-adjusted pricing. It surpasses prevalent replication strategies by far. Moreover, it allows to make the whole financial system more resilient through effective regulatory measures. 

The article is structured as follows. In the Section~\ref{sec:static}, we describe a prevalent replication methodology for retail banks. Furthermore, we specify similar use cases from other industries. Subsequently, we explain the key notion of Deep~ALM in the Section~\ref{sec:dalm}. A hint about the tremendous potential of Deep~ALM is provided in the Section~\ref{sec:drp} in the context of a prototype. While accounting for the regulatory liquidity constraints, a deep neural network adapts a non-trivial dynamic replication strategy for a runoff portfolio that outperforms static benchmark strategies conclusively. The last section elaborates on implications of Deep~ALM and future research. Common misunderstandings from the computational and regulatory viewpoint are going to be clarified.

\section{A Static Replication Scheme}\label{sec:static}

\subsection{The Treasury Case of a Retail Bank}

The business model of a retail bank relies on making profits from \emph{maturity transformations}. To this end, a substantial part of their uncertain liability term structure is re-invested. The liability side mainly consists of term and non-maturing deposits. Major positions on the asset side are corporate credits and mortgages. Residual components serve as liquidity buffers, fluctuation reserves and general risk-bearing capital. The inherent optimisation problem within this simplified business situation is already involved. Whereas most cash-flows originating from the liability side are stochastic, the bank is obliged to ensure a well-balanced liquidity management. Regarding short- and long-term liquidity, regulations require them to hold a \emph{liquidity coverage ratio} (LCR) and a \emph{net stable funding ratio} (NSFR) beyond $100\%$. In contrast, market pressure forces them to tighten the liquidity buffers in order to reach a decent return-on-equity. \emph{Asset-liability-committees} (ALCO) steer the balance sheet management through risk-adjusted pricing and hedging instruments. Due to the high complexity, many banks predominantly define and monitor static replication schemes. It is worth noticing that ALCOs nowadays often meet only once a month and supervise simplistic measures such as the modified duration and aspects of hedge accounting on loosely time-bucketed liabilities. Many opportunities are overseen, and many risks remain unhedged; e.g., see the introductory chapter of \cite{SDR19}. Crucial business decisions are inconsistent over time and rather reactionary than pro-active.

Table~\ref{tbl:static} illustrates the replication scheme of a generic retail bank. It is inspired from Section~4.2 in \cite{SDR19}. The notional allocated to fixed income currently amounts to EUR $100m$, which are rolled-over monthly subject to some static weights. The first line corresponds to a liquidity buffer that is held in non-interest bearing legal tender.

\begin{table}
\begin{tabular}{|r|r|r|r|r|}
\hline
\textbf{term in}&&\textbf{volume in}&\textbf{number of}&\textbf{volume in mEUR}\\
\textbf{months}&\textbf{weight in $\%$}&\textbf{mEUR}&\textbf{tranches}&\textbf{per tranche}\\
\hline
$0$&$10\%$&$10$&&\\
\hline
$1$&$15\%$&$15$&$1$&$15.000$\\
\hline
$3$&$15\%$&$15$&$3$&$5.000$\\
\hline
$6$&$15\%$&$15$&$6$&$2.500$\\
\hline
$12$&$15\%$&$15$&$12$&$1.250$\\
\hline
$60$&$15\%$&$15$&$60$&$0.250$\\
\hline
$120$&$15\%$&$15$&$120$&$0.125$\\
\hline
\multicolumn{1}{c|}{}&\textbf{100\%}&\textbf{100}&\textbf{202}&\textbf{24.125}\\
\hhline{~----}
\end{tabular}
\luft
\caption{Static investment scheme of a fixed income portfolio}\label{tbl:static}
\end{table}

Exemplarily, $15\%$ or EUR~$15m$ of the currently allocated capital are invested in fixed income instruments with a maturity of $1y$. This portion is again split into equal portions, such that roughly the same volume matures each month. In order to end up with a closed cycle, this refers to as maintaining $12$ tranches with a volume of EUR~$1.25m$ $(=1/12\times 15m)$. This redemption amount is re-invested again over a horizon of $1y$ at the latest market rates. The same roll-over procedure is conducted for all other maturities. This leaves us with $202$ rolling tranches in total. Under the current premises, $6$ tranches with a total volume of EUR~$24.125m$ mature each month and are re-invested consistently. The portfolio weights are only revised from time to time. They are typically the result of empirical and historical considerations. On the one hand, it is lucrative to increase the portion in longer terms as long-term yields are often but not always higher than short-term yields. On the other hand, if only a small portion is released each month, this leaves the bank with a significant interest and liquidity risk. This static scheme is typically accompanied with selected hedging instruments such that the sensitivities with respect to parallel shifts of the yield curve or the so-called \emph{modified durations} for the asset and liability side roughly coincide.

Certainly, dynamic strategies (including a restructuring of all tranches) are superior to their static counterparts regarding profit potential and resilience of the enterprise. However, dealing directly with $202$ tranches is often not feasible computationally. If we only add a simple dynamic feature and allow for arbitrary weights in the monthly re-investment process, this already introduces a considerable additional complexity from the analytical viewpoint. Even if one reduces the dimensionality of the yield curve dynamics, one faces an intricate nested optimisation exercise over all re-investment instances. Utilising traditional methods such as dynamic programming and accounting for a reasonable set of constraints is far from being trivial. Moreover, due to the necessary model simplifications, it is uncertain whether this additional complexity really pays off. It is therefore not surprising that many retail banks have been implementing a rather static replication scheme as described above. In the near future, this is likely to change once the novel and computationally much less intensive Deep ALM with a huge potential will be deployed.

\subsection{Similar Use Cases}

The notion of replication schemes may be applied analogously to other use cases. In the following, we describe three of them.

\subsubsection{Actuarial Perspective}

The situation described above can be transferred one-to-one to the situation of insurance companies. The aspects that change are completely different regulatory circumstances, the asset illiquidity of hedging instruments (the \emph{retrocession}) and the stochastic cash-flow generation as well as separate extreme risks on the liability side. Life insurance companies and pension funds must account for the exact product terms and the mortality characteristics of their policyholders. This includes, for instance, longevity and pandemic risks. Non-life insurance companies model the annual loss distribution and its unwinding over consecutive years.

\subsubsection{Procurement}

The producing industry is faced with a long and complex process chain. Several raw commodities need to be bought, delivered, stored, processed and designed into a consumer product. In order to control the inherent business risks and ensure profitability, a maturity transformation in the procurement process becomes inevitable. Even though the exact amount of the required materials is yet unknown, the company's exposure towards adverse price movements needs to be hedged accordingly. Despite the seemingly large daily trading volumes, any order comes with a price impact. Controlling the inventory is also worthwhile due to funding and storage cost. Therefore, slightly more involved than the treasury case described above, we have an ALM optimisation in the presence of illiquidity, storage cost, transaction cost and uncertainty.

\subsubsection{Hydroelectric Power Plants}

The technological advances regarding renewable energy and the deregulation are having a huge impact on electricity markets. Wind and solar energy make the electricity prices weather-dependent and more volatile. Still, some segments of future markets have remained comparatively illiquid. A peculiarity of the electricity market are its intricate price dynamics (including negative prices) and a wide range of more or less liquidly traded future contracts over varying delivery periods (e.g., intraday, day ahead, weekend, week, months, quarters, years). This makes the estimate of \emph{hourly price forward curves} (HPFC) that account for seasonality patterns (intraday, weekday, months), trends, holiday calendars and aggregated price information inferred from various future contracts (base, peak and off-peak) a very challenging exercise. Electricity producers may incur significant losses due to adverse contracts with locked-in tariffs and mis-timed hydropower production in reservoirs. This environment forces the suppliers to conduct a rigorous ALM, which comprises, for instance, a competitive pricing of fixed-for-floating contracts and daily optimised power production plans. The situation of delivering competitive yet still profitable margins in the presence of uncertainty is a situation banks have been accustomed to for decades (with a different commodity and different rate dynamics). Deep~ALM can be tailored to this very situation and provides exactly the technological solution to this delicate business problem. All one needs to establish is an appropriate scenario generator for the future markets and the HPFC. Many business-specific peculiarities can hardly be reflected by traditional optimisation techniques. As an illustration, a turbine cannot be turned on and off on an hourly basis. This simple-looking constraint renders the dynamic programming principle impossible. However, this constraint can be accounted for in Deep~ALM without further ado by incorporating suitable penalties.

\section{Deep ALM}\label{sec:dalm}

Beyond professional judgment, practical problems arising in ALM are often tackled either by linear programming or dynamic programming. The latter requires that the problem is divisible into simpler subproblems. For non-linear and high-dimensional cases, one typically exploits Monte-Carlo techniques in order to derive or validate strategies empirically along a handful of criteria. As a matter of fact, the level of sophistication for replicating strategies remains fairly limited due to the high intricacy and the lack of an adequate constructional approach. Deep~ALM makes all these impediments obsolete. One directly implements the arbitrarily complex rule book of the use case and lets a very smart and non-forgetful artificial financial agent gain the experience of many thousand years in a couple of hours. By incentivising the desired behaviour and stimulating a swift learning process, one reaches superhuman level in due course.

\begin{figure}
\includegraphics[page=1,trim={4.5cm 16.6cm 4.5cm 4.1cm},clip,width=\textwidth]{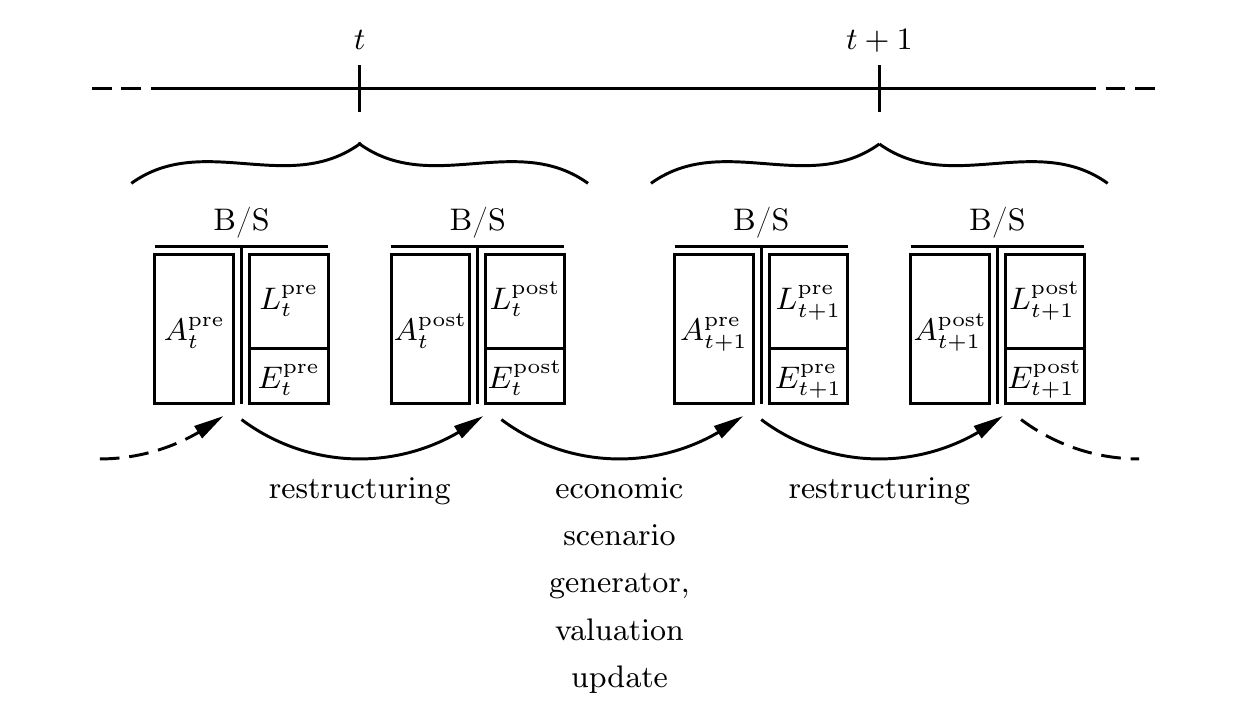}
\caption{Balance sheet roll-forward}
\label{fig:bs}
\end{figure}

We consider a balance sheet evolution of a retail bank over some time grid $\{0,1,2,\hdots,N\}$ in hours, days, weeks or months. The roll-forward is an alternating process between interventions and stochastic updates over time; see Figure~\ref{fig:bs}. At each time instance $t$, an artificial asset-liability-manager, subsequently denoted the \quote{artificial financial agent} (AFA), assesses the re-investment of matured products as well as the restructuring of the current investment portfolio. This involves various asset classes, not just fixed income instruments. Correspondingly, the AFA is active in both the \emph{primary market} of newly issued instruments and the \emph{secondary market} of previously issued and circulating instruments. There may be further eligible transformations of this initial state with transaction cost, e.g., granting further credits or building additional liquidity buffers. The balance sheet components $A_t$ (assets), $L_t$ (liabilities) and $E_t$ (equity) come with the superscripts \quote{pre} and \quote{post}, which refer to as before and after the balance sheet restructuring respectively. Preceding interventions, an auxiliary step for the calculation of taxes may be necessary. Subsequently, the economic scenario generator performs a stochastic roll-forward of the balance sheet. All accounts are updated to the latest circumstances and macro-economic factors, e.g., certain products pay off cash amounts such as coupons or dividends, certain products need to be written off due to the bankruptcy of the referenced entity, clients withdraw certain portions of their deposits, etc. It needs to be noted that the described situation is kept simplistic for illustrative purposes. It is the main target of future research work to reach an acceptable level of sophistication in order to make the technology eligible for real world ALM.

\begin{figure}
\includegraphics[page=2,trim={4.5cm 18.9cm 4.5cm 4.1cm},clip,width=\textwidth]{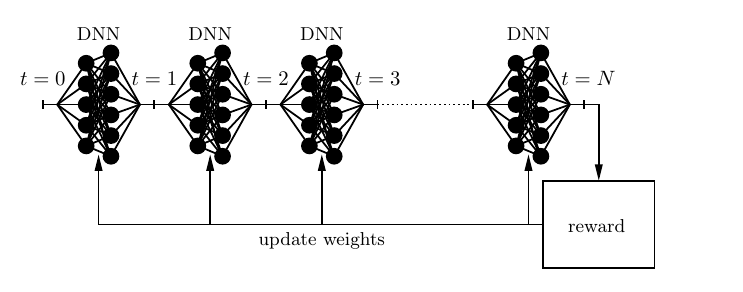}
\caption{Deep neural network architecture}
\label{fig:reinf}
\end{figure}

The superiority of the dynamic replication scheme requires that the trained AFA can perfectly trade-off benefits from a restructured portfolio and the incurred transaction cost of the restructuring. Furthermore, the trained AFA must impeccably weigh up different objectives such as complying with regulatory constraints, earning a risk-adjusted spread, maintaining an adequate level of liquidity, keeping sufficient margins for adverse scenarios, etc. The fundamental idea is to parameterise the decision process at each time instance $t$ through a deep neural network (DNN) and to incrementally improve its performance through techniques inspired from machine learning. By concatenating these deep neural networks along the time axis, one ends up at a holistic dynamic replication strategy that easily deals with non-stationary environments. The learning process incentivises the maximisation of an adequately chosen reward function by incrementally updating the weights of the feedforward networks through \emph{backpropagation}; see Figure~\ref{fig:reinf}. This idea allows to tame the curse of dimension. These new possibilities of Deep~ALM allow for a more frequent supervision while accounting for arbitrarily many side constraints and complicated market frictions such as illiquidity and offsetting effects of transaction cost in hedges. Thus, the full power of dynamic strategies can be deployed.

Traditional numerical schemes describe an algorithm either in order to solve an equation or in order to optimise a certain functional. Typically, algorithms are only accepted by the mathematical community if they come with two supplements, namely a proof that the recipe terminates with a certain accuracy at the desired level, on the one hand, and a statement about the rate-of-convergence, on the other hand. In the context of Deep~ALM, we coin the notion of \emph{convincing strategies} because we are likely not able to prove that the randomly initialised deep neural network will end up at some optimal replication strategy. In fact, we intend to derive non-trivial but still realistic dynamic replication strategies within a reasonable time that significantly outperform those strategies currently used in the financial industry. From the mathematical viewpoint, Deep~ALM entails a delicate paradigm shift. The results will be derived empirically based on scenarios representing the experience of many thousand years. By repeating the learning process a few times, we will corroborate the robustness of the approach. Furthermore, we will validate and challenge the performance of the deep neural network on unseen testing data. If the gradient descent algorithm can only further reduce the training loss with some overfitting strategy at the cost of the validation loss, we have ended up at a local optimiser. Even though we will not be able to prove global optimality, the strategy might still reach superhuman level and, thus, be more than convincing. 

\section{Empirical Study on a Stylised Case}\label{sec:drp}

\subsection{The Runoff Case}

We aim at optimising the replication strategy when unwinding a runoff portfolio over ten years in the context of an equity index and a large fixed income universe. At every time instance, roughly 200 bonds are active. This prototype demonstrates the feasibility of Deep~ALM in a stylised case. In the end, every balance sheet roll-forward is a superposition of runoff portfolios. Furthermore, we give an indication about the potential of dynamic replication strategies.

The liabilities are backed by a static legacy investment portfolio. Roughly two thirds of the assets are invested in the fixed income market and $10\%$ are invested in equities. The remainder is held in cash. Every month, six new bonds are issued in the primary market with the face amount of $100$ monetary units and the maturities $1m$, $3m$, $6m$, $1y$, $5y$ and $10y$. The first three series only pay a coupon at the final redemption date, the other three pay semi-annual coupons. At issuance, all bonds trade at par. The legacy investment portfolio has attributed equal portions to all of the six bond series and has conducted an equally distributed roll-over strategy based on historical data from the European Central Bank (ECB). Future yield curve scenarios are simulated consistent with a principal component analysis. For simplicity, we assume that there is no secondary market. Hence, bonds cannot be sold on in the market and must be held to maturity. Furthermore, we preclude credit defaults. Regarding equities, we generated scenarios based on a discretised geometric Brownian motion. Changing the position in equities involves proportional transaction cost. Due to legal requirements, at least $10\%$ of the asset portfolio must be held in cash; this is just an arbitrary choice in order to incorporate a regulatory constraint. Any discrepancy is penalised monthly with a high interest rate. The initial balance hosts assets worth $100$ monetary units. The term structure of liabilities with the same face amount is spread across the time grid according to some beta distribution.

\subsection{The Implementation Strategy}

A well-established approach to tackle quantitative finance problems with deep reinforcement learning is to follow the standard routine inspired from games. A game is a stochastic environment in which a player can act according to a set of rules and can influence the course of events from start to finish to a certain degree. The game terminates after a prespecified number of rounds. A rule book includes the model drivers and their interdependence, the constraints and the penalties for constraint violations, the control variables and the objective function. Notably, we actually do not solve the game for all initial states nor do we assume any Markovian structure. We implemented the 'game' according to the rule book specified below in a Jupyter Notebook such that the random outcome of any strategy could be simulated arbitrarily often. To this end, we defined an \emph{economic scenario generator} that hosted the yield curve dynamics and spot prices for equities. Having this environment in place, one can test the performance of any na\"ive or traditionally derived replication portfolio. The step of translating the rule book into model dynamics is crucial for understanding the game, for ensuring a well-posed optimisation problem and for streamlining the logic of the subsequent tensorisation. The evolution of $\R^N$-valued quantities can be inferred from historical time series with a principal component analysis as proposed by Andres Hernandez in \cite{H16}. Subsequently, the game is embedded in a deep neural network graph architecture using TensorFlow such that an AFA can run many games in parallel and adaptively improve its performance. The graph constitutes of many placeholders and formalises the logic of the game. The first decision process of the AFA is managed through the randomly initialised weights of a neural network with a handful of hidden layers. Its input layer is a parameterisation of the initial state, concerning both the balance sheet structure and the market quotes, and the output layer characterises the eligible balance sheet transformation. The balance sheet roll-forward is accomplished by connecting the output of the neural network with a whole bunch of efficient tensor operations reflecting the balance sheet restructuring, new inputs for the non-anticipative scenario updates and the input layer of the next decision process. Iterating this process yields the desired deep neural network.

\subsection{Term Structures}

We are given a discrete time grid $\mathbb{T}_0=\{0,1,2,\hdots,N\}$ in months, where $N=120$ refers to as a planning horizon of ten years. The initial term structure of liabilities with the face amount of $100$ monetary units is spread across the time grid according to a beta distribution with the parameters $a=1.5$ und $b=2.5$, i.e., if $F_{a,b}$ denotes the corresponding cumulative distribution function, then the cash-flow
\[L^{(t)}:=\Big(F_{a,b}\big(t/N\big)-F_{a,b}\big((t-1)/N\big)\Big)\times 100\]
becomes due in month $t\in\mathbb{T}:=\mathbb{T}_0\setminus\{0\}$. We briefly write $L:=\big(L^{(1)},L^{(2)},\hdots,L^{(N)}\big)$ for the collection of all payables. At some fixed time instance, any active bond is characterised through its future cash-flows $B=\big(B^{(1)},B^{(2)},\hdots,B^{(N)}\big)\in\mathbb{R}^N$. If $Y=\big(Y^{(1)},Y^{(2)},\hdots,Y^{(N)}\big)\in\mathbb{R}^N$ denotes the current yield curve with the yield $Y^{(T)}$ for the maturity $T/12$ in years and $D=\big(D^{(1)},D^{(2)},\hdots,D^{(N)}\big)\in\mathbb{R}^N$ with $D^{(T)}=e^{-T/12\cdot Y^{(T)}}$ for all $T\in\mathbb{T}$ the consistent discount factors, then the value of $B$ is simply $V(B)=\langle D,B\rangle$, where $\langle\cdot,\cdot\rangle$ stands for the standard inner product. The coupons are chosen such that each bond trades at par. The cash-flow generation at issuance is straightforward. It corresponds to solving a linear equation. Exemplarily, a bond with a maturity of $5y$ and semi-annual coupons carries an annualised coupon rate
\[c=\frac{12}{6}\cdot\frac{1-D^{(60)}}{\displaystyle\sum_{k=1}^{10}D^{(6\cdot k)}}.\]
Therefore, it holds
\[B^{(6\cdot k)}=\dfrac{c}{2}\cdot 100\quad\text{for all }k\in\{1,2,\hdots,9\},\qquad B^{(60)}=\bigg(1+\frac{c}{2}\bigg)\cdot 100,\]
and all other components are zero. When going from one time instance to the next, $B^{(1)}$ is paid off and one simply applies to the scheme $B$ the linear update operator
\[U=\left(\begin{matrix}0&1&&&0\\&0&1&\\&&\ddots&\ddots\\&&&0&1\\0&&&&0\end{matrix}\right)\in\mathbb{R}^{N\times N}.\]
A $k$-fold application of the update operator is denoted by $U^k$. We utilise data from the European Central Bank (ECB) in order to model the historical and future yield curves. ECB publishes for each working day the Svensson parameters $\beta_0$, $\beta_1$, $\beta_2$, $\beta_3$, $\tau_1$ and $\tau_2$ of its yield curve. The corresponding yield curve is the vector $Y=\big(Y^{(1)},Y^{(2)},\hdots,Y^{(N)}\big)\in\mathbb{R}^N$ with
\[Y^{(T)}=\beta_0+\beta_1\frac{1-e^{-T/(12\tau_1)}}{T/(12\tau_1)}+\beta_2\frac{1-e^{-T/(12\tau_1)}}{T/(12\tau_1)-e^{-T/(12\tau_1)}}+\beta_3\frac{1-e^{-T/(12\tau_2})}{T/(12\tau_2)-e^{-T/(12\tau_2)}}.\]
We fix some historical time series in $\mathbb{R}^N$ and conduct a principal component analysis. To this end, we assume stationary daily yield curve increments $\Delta X$ with expected value $\mathbb{E}[\Delta X]\equiv\mu\in\mathbb{R}^N$ and covariance matrix $cov(\Delta X)\equiv Q\in\mathbb{R}^{N\times N}$. We spectrally decompose $Q=\Lambda L\Lambda^\top$ into the normalised eigenvectors $\Lambda\in\mathbb{R}^{N\times N}$ and the ordered eigenvalues $L=\text{diag}\{\lambda^{(1)},\lambda^{(2)},\hdots,\lambda^{(N)}\}$ with $\lambda^{(1)}\geq\lambda^{(2)}\geq\hdots\geq\lambda^{(N)}$. The last historical yield curve is the day when we acquire the runoff portfolio. The stochastic yield curve increments from one month to the next are consistently sampled according to
\[\Delta Y=22\cdot\mu+\sum_{k=1}^nZ_k\Lambda_{\cdot k},\]
where we choose $n=3$ and $Z_k\sim\mathcal{N}\big(0,{\sigma_k}^2\big)$ for $\sigma_k=\sqrt{22\cdot\lambda^{(k)}}$. The factor $22$ accounts for the number of trading days per month. If $i$ denotes the maturity in number of months, the legacy bond portfolio has been investing each month a face amount of $15/i$ monetary units in that particular bond series for $i\in I:=\{1,3,6,12,60,120\}$. This is consistent with the replication scheme from Table~\ref{tbl:static}.

\subsection{The Equity Market}

We assume that the stochastic process $S=(S_t)_{t\in\mathbb{T}_0}$ follows a discretised geometric Brownian motion, i.e., it holds for all $t\in\mathbb{T}$
\[\log\frac{S_t}{S_{t-1}}\sim\mathcal{N}\Bigg(\bigg(m-\frac{1}{2}s^2\bigg)\cdot\frac{1}{12},\frac{s^2}{12}\Bigg)\]
with initial level $S_0=100$, drift $m=5\%$ and volatility $s=18\%$. Changing the position in equities involves proportional transaction cost $\kappa=0.50\%$.

\subsection{The Balance Sheet}

We proceed iteratively. Let $Y_t\in\R^N$ denote the current yield curve for $t\in\mathbb{T}_0\setminus\{N\}$ and $D_t$ consistent discount factors. Before restructuring, the left hand side of the balance sheet with a total present value of $A_t^\pre$ consists of three additive components: the held cash $C_t^\pre$, the legacy bond portfolio with the value $V_t^\pre$, and the equity position worth $G_t^\pre=\Delta_t^\pre\cdot S_t$. The right hand side of the balance sheet consists of the liabilities
\[L_t^\pre=\langle D_t,U^tL\rangle\]
and the residue $E_t^\pre=A_t^\pre-L_t^\pre$.

\subsection{The Deep Neural Network Architecture}

For any $t\in\mathbb{T}_0\setminus\{N\}$, we consider a feedforward neural network
\[F_t=\big(\phi\circ W_t^{(2)}\big)\circ\big(\phi\circ W_t^{(1)}\big)\circ\big(\phi\circ W_t^{(0)}\big)\]
with some affine functions
\[W_t^{(0)}:\mathbb{R}^{10}\longrightarrow\mathbb{R}^{15}, W_t^{(1)}:\mathbb{R}^{15}\longrightarrow\mathbb{R}^{15}, W_t^{(2)}:\mathbb{R}^{15}\longrightarrow\mathbb{R}^{7}.\]
 and the ReLU activation function $\phi(x)=\max\{x,0\}$. The input layer consists of the \emph{leverage ratio} $E_t^\pre/A_t^\pre$, the \emph{liquidity ratio} $C_t^\pre/A_t^\pre$, the \emph{risk portion} $G_t^\pre/A_t^\pre$, the holding in equities $\Delta_t^\pre$ and the yields $Y_t^{(i)}$ for $i\in I$ of the latest bond series. The output layer reveals the outcome of the restructuring at the time instance $t$. The first six components represent the holdings $h_t^{(i)}$ for $i\in I$ in the latest bond series, the last component represents the holding $\Delta_t^\post$ in equities. Note that the activation function implies long-only investments.
 
\subsection{The Restructuring}

The restructuring implicates the updated balance sheet items
\begin{align*}
C_t^\post&=C_t^\pre-100\cdot\sum_{i\in I}h_t^{(i)}-\Big(\big(\Delta_t^\post-\Delta_t^\pre\big)+\kappa\cdot\big|\Delta_t^\post-\Delta_t^\pre\big|\Big)\cdot S_t,\\
V_t^\post&=V_t^\pre+100\cdot\sum_{i\in I}h_t^{(i)},\\
G_t^\post&=\Delta_t^\post\cdot S_t,\\
A_t^\post&=C_t^\post+V_t^\post+G_t^\post,\\
L_t^\post&=L_t^\pre,\\
E_t^\post&=A_t^\post-L_t^\post.
\end{align*}

\subsection{The Roll-Forward}

Let $B_t^\post\in\R^N$ denote the aggregated future cash-flows of the whole fixed income portfolio after restructuring at time $t$. Furthermore, let $\pi^{(1)}:\R^N\longrightarrow\R$ denote the projection onto the first component. If one does not adhere to the regulatory constraint of holding at least $10\%$ of the assets in cash, one is penalised with liquidity cost of $p=24\%$ per annum on the discrepancy. Once the yield curve has been updated stochastically to the state $Y_{t+1}$ with consistent discount factors $D_{t+1}$ and the equity index has been updated stochastically to the state $S_{t+1}$, this leaves us with the roll-forward
\begin{align*}
C_{t+1}^\pre&=C_t^\post+\pi^{(1)}\big(B_t^\post\big)-\pi^{(1)}\big(U^tL\big)-\frac{p}{12}\cdot\max\big\{10\%\cdot A_t^\post-C_t^\post,0\big\},\\
V_{t+1}^\pre&=\langle D_{t+1},UB_t^\post\rangle,\\
G_{t+1}^\pre&=\Delta_t^\post\cdot S_{t+1}=\Delta_{t+1}^\pre\cdot S_{t+1},\\
A_{t+1}^\pre&=C_{t+1}^\pre+V_{t+1}^\pre+G_{t+1}^\pre,\\
L_{t+1}^\pre&=\langle D_{t+1},U^{t+1}L\rangle,\\
E_{t+1}^\pre&=A_{t+1}^\pre-L_{t+1}^\pre.
\end{align*}

\subsection{Objective}

For different maturities $T\in\mathbb{T}$, a common approach in mathematical finance is to maximise $\mathbb{E}\big[u\big(E_T^\text{post}/E_0^\text{pre}\big)\big]$, where
\[u(x)=\begin{cases}\dfrac{x^{1-\gamma}-1}{1-\gamma}&,\text{ if }\gamma\neq1\\\log{x}&,\text{ if }\gamma=1\end{cases}\]
denotes the iso-elastic utility function with constant relative risk aversion $\gamma\geq 0$. This is not directly applicable in our case since we cannot prevent $E_T^\text{post}$ from becoming negative. Instead, we aim at maximising
\[\mathbb{E}\bigg[u\bigg(\Big(\varepsilon+\phi\big(E_T^\text{post}\big)\Big)/E_0^\text{pre}\bigg)\bigg]\]
for a small constant $0<\varepsilon\ll 1$. Provided that the final equity distribution is positive, the case $\gamma=1$ corresponds to the quest for the \emph{growth optimal portfolio} that maximises the expected log-return; see \cite{PH06}. An intriguing alternative is quadratic hedging, where one aims at minimising
\[\mathbb{E}\bigg[\Big(E_T^\text{post}-(1+r)^{T/12}E_0^\text{pre}\Big)^2\bigg]\]
for some annualised target return-on-equity $r>0$. Either case leads to convincing strategies.

\subsection{Results}

\begin{figure}
\includegraphics[page=3,trim={4.25cm 15.4cm 4.75cm 4.1cm},clip,width=\textwidth]{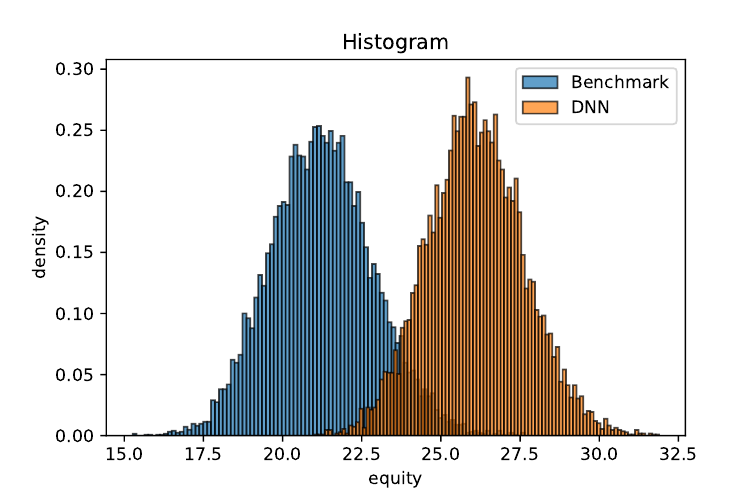}
\caption{Line-up between the benchmark and the trained AFA}
\label{fig:python}
\end{figure}

The optimisation is only non-trivial for non-negative yield curves. If the mid-term yields are consistently negative (as they have been in the EUR area for quite some years), holding cash is superior to the considered fixed income investments (within the simplified premises specified above). As starting point, we choose the ECB yield curve as per December~31, 2007. In that case, the initial value of the liabilities amount to $L_0^\pre=86.0$ monetary units. As outlined above, the exercise of optimising the equity distribution after the settlement of all liabilities is extremely challenging with conventional methods. The investment universe consists at each time instance of $204$ active assets, namely the fixed income universe, an equity index and legal tender. We propose a simple benchmark strategy that allocates the available cash beyond $10\%$ of $A_t^\text{pre}$ to bonds with a maturity of $1m$, provided that the coupons are positive. Equities are divested linearly over time. Moreover, we train an AFA on $10\,000$ unwinding scenarios following the ideas of Deep~ALM. An until only recently inaccessible dynamic strategy outperforms the benchmark strategy conclusively after a short learning process of roughly $10$~minutes. Using another $10\,000$ previously unseen validation scenarios, the following chart illustrates the final equity distributions after ten years for a classical static strategy in blue and for the dynamic strategy in orange. The dynamic strategy does not involve extreme risk taking. It simply unveils hidden opportunities. The systematic excess return-on-equity with respect to the benchmark is beyond $2\%$ per annum; see Figure~\ref{fig:python}.

\section{Outlook and Further Research}\label{sec:outlook}

\subsection{Potential of Deep~ALM}

The real potential can hardly be foreseen. One achievement will lead to another and trigger further accomplishments. Despite the optimism, the technique is not mature enough and requires an adequate level of research. Market participants across all use cases, who we spoke with, like the intriguing idea of Deep~ALM. However, only a few dare being the first movers in this precompetitive phase. ALM is at the heart of risk management for any enterprise in the financial industry. In the light of the new opportunities, it is not acceptable how simplistic and ineffective prevalent replication schemes are to this day. All of us are directly affected by unhedged risks of bank deposits, insurance policies and pension funds. This can be illustrated exemplarily with Finland whose banking sector caused a devastating financial crisis in the early 1990s. In the aftermath, official unemployment rates escalated to roughly $18\%$. Deep~ALM offers a powerful framework that supports well-balanced risk-taking and unprecedented risk-adjusted pricing. Thus, its rigorous application may prevent a financial crisis sooner or later. Likewise, we are all exposed to the climate change. Hence, it is becoming increasingly important to spend resources wisely. Deep~ALM promotes this attitude with significant efficiency gains in the procurement of commodities and the hydropower production. These are promising prospects for Deep~ALM indeed.

\subsection{Necessity of Further Research}

Some people argue that machine learning will facilitate the enterprise risk management, make it more effective and make some of the current tasks even redundant. We agree that the new opportunities will allow for more consistent and sound risk taking. However, we are convinced that legitimate applications of Deep~ALM will entail a lot of additional work since the previously inaccessible analyses do not come for free. Moreover, they will raise a huge portion of new economic challenges and obligations. Thus, it is essential that the research community prepares a solid ground for this technological transformation.

\subsection{Feasibility}

Optimising a whole term structure of assets and liabilities by utilising techniques inspired from deep reinforcement learning is a novel approach and its feasibility needs to be demonstrated. In the light of recent advances, we are optimistic that this goal can be achieved in due course. Nonetheless, this challenge opens a great field of research questions from the mathematical, computational and economic viewpoints. Regarding feasibility, it is an important concern that we do not just want to feed an utterly complex mathematical system into a powerful supercomputer and see whether we get something out. In contrast, we aim at formulating and optimising dynamic replication schemes that meet the requirement of industrial applicability, on the one hand, and operability on a moderately enhanced computer, on the other hand. Particularly, this involves an assessment whether the behaviour of the trained AFA is interpretable, whether it follows sound economic reasoning and whether the non-anticipative dynamic replication strategies fulfil the basic requirement of practical viability. Solving a high-dimensional utility maximisation problem that entails unrealistic leverage or arbitrary rebalancing frequencies is futile for real world applications. For this purpose, Deep~ALM must trade off many objectives such as cost and benefit of liquidity, drawdown constraints, moderate leverage ratios, profit expectation, regulatory interventions, risk appetite, short sale restrictions, transaction cost, etc. Most aspects can be accounted for in the model by means of penalties that incentivise the AFA’s behaviour accordingly. Others such as drawdown constraints and profit expectation typically enter the target function of the optimisation. Finding the right balance between all the different goals without adversely affecting the robustness of the learning process entails some engineering work. However, once a convincing and stable economic model choice has been established, this has a tremendous impact on the economic research field. Previously inaccessible price tags to crucial economic notions such as, for instance, cost of liquidity for a retail bank can be evaluated quantitatively. All one needs to do is to line up performance measures of trained AFAs when short sale restrictions of the cash position are enforced and dropped respectively.

\subsection{Goal-based Investing}

A closely related subject to Deep~ALM is goal-based investing; e.g., see \cite{Br99} by Sid Browne for a neat mathematical treatment. One aims at maximising the probability to reach a certain investment goal at a fixed maturity. From the theoretical viewpoint with continuous rebalancing, this is equivalent to replicating a volume-adjusted set of digital options (unless the target return is smaller or equal than the risk-free rate). Despite being very intuitive, this result is only of limited use for practical applications. Discrete rebalancing and the payoff-discontinuity at the strike lead, as one gets closer and closer to maturity, to an inherent shortfall risk way beyond the invested capital. It needs to be investigated whether Browne’s setting can be modified accordingly for real world applications. This involves more realistic dynamics of the economic scenario generator, exogenous income (as inspired by Section~7 of \cite{Br99}), maximal drawdown constraints and a backtesting with historical time series.

\subsection{Generalisation}

If we take one step further and look at ALM as an abstract framework, Deep~ALM may contribute crucially to urgent challenges of our society such as combating the climate change or dealing with pandemics. It trains a player to reach a certain goal while minimising the involved cost. Regarding the climate change, there is a complex trade-off in implementing protection measures, incentivising behavioural changes, transforming the energy mix, etc. Regarding COVID-19, rigorous emergency policies were enforced throughout the world in order to protect individual and public health. These measures involved extreme cost such as temporary collapses of certain industries, closed schools, restricted mobility, etc. These are again high-dimensional problems, which cannot be tackled with classical methods. Deep~ALM is a first step towards these more difficult challenges.

\subsection{Impediments}

A challenge will be to overcome the common misunderstanding that deep learning requires a lot of historical data (that banks or commodity companies usually do not have), that deep learning is an incomprehensible black box, and that regulators will never approve of Deep~ALM. Firstly, the above approach does not require any historical data. All one needs are model dynamics of ordinary scenarios and stress scenarios. The training data can be fully built on these premises synthetically. Secondly, the performance of deep neural networks can be validated without further ado along an arbitrarily large set of scenarios. To our mind, this is one of the most striking features of Deep~ALM (which is not satisfied by traditional ALM models). Traditional hybrid pricing and risk management models entailed an intensive validation process. This intensive work becomes obsolete and can be invested into stronger risk management platforms instead. Thirdly and lastly, as long as one provides the regulators with convincing arguments regarding the accuracy and resilience of the approach (e.g., in terms of a strong validation and model risk management framework), they will never object to use deep learning.

\subsection{Regulatory Perspective}

The complex regulatory system in place is supposed to make the financial system less fragile. However, designing and implementing effective policies for the financial industry, which do no promote wrong incentives and which are not (too) detrimental to economic growth, is very demanding. This is highlighted exemplarily in the IMF working paper \cite{SCK03}. An illustration that certain measures do not necessarily encourage the desired behaviour can be found in \cite{GLD16} by Martin Grossmann, Markus Lang and Helmut Dietl. Initially triggered by the bankruptcy of Herstatt Bank (1974), the Basel Committee on Banking Supervision has established a framework that specifies capital requirements for banks. It was revised heavily twice since its publication in 1988. The current framework, also known as \quote{Basel~III} (see \cite{BIS11}), was devised in the aftermath of the 2007/2008 credit crunch. Amongst others, it proposed requirements for the LCR and the NSFR. The impact of these liquidity rules was empirically assessed in \cite{DHW14} by Andreas Dietrich, Kurt Hess and Gabrielle Wanzenried. Due to limited historical data and the intricate complexity between the plethora of influencing factors, the true impact of the new regime can hardly be isolated. Deep~ALM opens an extremely appealing new way of appraising the effectiveness of regulatory measures. To this end, one investigates how a very smart and experienced AFA changes her behaviour when certain regulatory measures are enabled or disabled respectively. This approach allows to truly identify the impact of supervisory interventions and the compatibility amongst each other. These quantitative studies will be a promising supplement of Deep~ALM.

\subsection{Model Risk Management}

The supervisory letter SR 11-7 of the U.S.~Federal Reserve System (\quote{Fed}) has become a standard for considerations of \emph{model risk management}; see \cite{Fed11}. Controlling the inherent model risk and limitations will be a key challenge before Deep~ALM may be exploited productively. While the performance of proposed strategies can be checked instantaneously for thousands of scenarios within the model, an assessment of the discrepancy between the Deep~ALM model and the real world is a crucial research topic. Once a robust Deep~ALM learning environment has been established, these aspects can be explored by means of various experiments. For instance, one can analyse the sensitivities of the acquired replication strategies with respect to the model assumptions or the performance of those in the presence of parameter uncertainties. These uncertainties may concern both the assumed states (e.g., the term structure of deposit outflows) and the concealed scenario generation for both the training and the validation (e.g., the drift assumptions of equities).

\subsection{Risk Policy Perspective}

Another intriguing question is the quest for locally stable strategic asset allocations. Typically, the trained neural network will reshuffle the balance sheet structure right from the beginning. Characterising \emph{no-action regions} corresponds to locally optimal initial states. Concerning the model risk management, revealing the link between different model assumptions and the internal risk policies, on the one hand, and the no-action regions, on the other hand, will be powerful tool for risk committees.

\bibliographystyle{amsplain}

\end{document}